\documentclass{article}

     \PassOptionsToPackage{numbers, compress}{natbib}
     \usepackage[final]{neurips_2019}

\usepackage[utf8]{inputenc} 
\usepackage[T1]{fontenc}    
\usepackage{hyperref}       
\usepackage{url}            
\usepackage{booktabs}       
\usepackage{amsfonts}       
\usepackage{nicefrac}       
\usepackage{microtype}      
\usepackage{subfigure}      
\usepackage{caption}
\usepackage{graphicx}
\usepackage{amsmath}
\usepackage{booktabs, tabularx}
\usepackage{float}

\title{Flatsomatic: A Method for Compression of Somatic Mutation Profiles in Cancer}

\author{
  Geoffroy Dubourg-Felonneau$^1$, Yasmeen Kussad$^{1,2}$, Dominic Kirkham$^{1,3}$, \\
  \textbf{John W Cassidy$^1$, Nirmesh Patel$^1$, Harry W Clifford$^1$} \\
  \\
  \and
  $^1$Cambridge Cancer Genomics \\
  Cambridge, UK \\
  www.ccg.ai
  \and
  $^2$University of Lancaster \\
  Lancaster, UK \\
  \and
  $^3$University of Cambridge \\
  Cambridge, UK \\
}

\begin{document}

\maketitle

\begin{abstract}

As part of the decision-making process in cancer therapy selection, oncologists stratify patients into broad clinical groups. At present, the major indicator for stratification is the site of origin of the primary tumor. However, with the widespread adoption of next generation sequencing technologies, our reference for stratification is becoming more complex as we are beginning to understand how each tumor is unique on the genetic level \cite{precisionpanc}. The outcome of this is an approach dubbed precision oncology, which involves the process of identifying genomic features driving an individual tumor and designing a personalized therapeutic strategy in response \cite{precisiononcology}. This presents a classification problem that is well suited to supervised machine learning algorithms, although due to the high complexity and dimensionality of such genomic data, applying models directly on the raw data can be difficult \cite{prior}. Common methods for reducing the dimensionality include incorporation of domain expertise to select features with a high likelihood of impact, for example driver genes\cite{drivers} or cell signaling pathways\cite{pathways}. The problem with this approach is it does not allow for discoveries of novel sources of signal within the data.

\smallskip

In this study, we present Flatsomatic - a Variational Auto Encoder (VAE) optimized to compress somatic mutations that allow for unbiased data compression whilst maintaining the signal. We compared two different neural network architectures for the VAE: Multilayer Perceptron (MLP) and bidirectional LSTM. The somatic profiles we used to train our models consisted of 8,062 Pan-Cancer patients from The Cancer Genome Atlas\cite{TCGA} and 989 cell lines from the COSMIC cell line project\cite{CCLP}. The profiles for each patient were represented by the genomic loci where somatic mutations occurred and, to reduce sparsity, the locations with a frequency <5 were removed. We enhanced the VAE performance by changing its evidence lower bound, and devised an F1-score based loss showing that it helps the VAE learn better than with binary cross-entropy. We also employed beta-VAE to weight the variational regularisation term in the loss function and showed the best performance through a preliminary function to increase the weight of the regularisation term with each epoch. We assessed the reconstruction ability of the VAE using the micro F1-score metric and showed that our best performing model was a 2-layer deep MLP VAE. Our analysis also showed that the size of the latent space did not have a significant effect on the VAE learning ability. We compared the Flatsomatic embeddings created to a lower dimension version of the data from principal component analysis, showing superior performance of Flatsomatic, and performed K-means clustering on both datasets to draw comparisons to known cancer types of each profile. Finally, we present results that confirm that the Flatsomatic representations of 64 dimensions maintain the same predictive power as the original 8,298 dimensions vector, through prediction of drug response.

\smallskip

Our work shows a great potential for the use of VAEs in creating lower dimension representations of somatic mutation profiles. We believe that work such as this is a highly important step in the use of somatic mutation data for machine learning applications, and we hope this will help future researchers in bringing data-driven models into the field of precision oncology.

\end{abstract}

\section{Introduction}
Analysis of somatic mutation profiles from cancer patients is essential in the development of cancer research. However, the low frequency of most mutations and the varying rates of mutations across patients makes the data noisy, sparse and extremely high dimensional. Such data is challenging to statistically analyze as well as difficult to use in classification problems, for clustering, visualisation or for learning useful information. Thus, the creation of low dimensional representations of somatic mutation profiles that hold useful information about the DNA of cancer cells will facilitate the use of such data in applications that will progress precision medicine. In this paper, variational autoencoders (VAEs) \cite{RN27} were used to create latent representations of somatic profiles.

\section{Methods}
The process of building the VAEs to compress somatic profiles was comprised of exploring different neural network architectures and optimizing them to enhance their performance. Several changes to the loss function of the VAE were explored \cite{RN15} in order to learn more useful representations. All models were implemented using Keras library \cite{RN40} with TensorFlow backend. 

The somatic profiles used to train our models were comprised of 8062 Pan-Cancer patients from The Cancer Genome Atlas (TCGA)\cite{RN30}, and  989 cell lines from the COSMIC cell line project (CCLP). The profiles for each patient are represented by the positions in the genome where the samples have a somatic mutation. To pre-process the data, binary vectors of equal lengths were created for each patient with ones marking the presence of a somatic mutation in a particular somatic mutation location. To reduce the sparsity, the mutation locations with a frequency of less than 5 were removed.

\subsection{Multi Layer Perceptron based models}
The first architecture explored was a feed-forward network also known as multilayer perceptron (MLP). A two-layer deep network in the encoder/decoder of the VAE were built with a batch normalization \cite{RN39} layer after each layer. Different combinations of the number of units in each layer were attempted and the effect of changing the size of the latent space was studied. A leaky ReLU \cite{LRELU} activation was used in the encoder layers, and a regular ReLU was used in the decoder layers except for the final layer in the decoder which employed a sigmoid function. A dropout layer was added after the first layer in the encoder/decoder, and  L1 regularization was used with each dense layer. We optimized with RMSprop, and the models were trained with a batch size of 128 for 100 epochs.

    \begin{figure}[H]
        \centering\includegraphics[width=.75\linewidth]{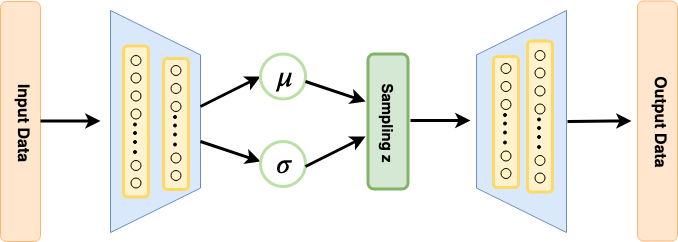}
        \centering\caption{General architecture of MLP-VAE. The orange layers are the input and output data. The light blue trapezoids represent the encoder and decoder layers with hidden layers. The green part in the middle represents the latent space where the parameters outputted from the encoder are used to sample z and create the embeddings. The embeddings are then inputted into the decoder to reconstruct the data.}
    \end{figure}

\subsection{Bi-directional LSTM}
The second architecture attempted was a bidirectional LSTM \cite{Schuster1997BidirectionalRN} in the hopes that it would reduce the significance of the order of mutations, especially those that are close to the end of the input sequence. Although the order of mutations in a somatic profile are not important, we expect the occurrence of certain mutations or a group of mutations together to hold important information, as a result this architecture was deemed the most suitable to learn deeper patterns in the data. The number of units in the LSTM was explored (1024, 512, 256, or 128) as well as the size of the latent space (128, 64, 32, or 8).
\smallskip

    \begin{figure}[H]
        \centering\includegraphics[width=.75\linewidth]{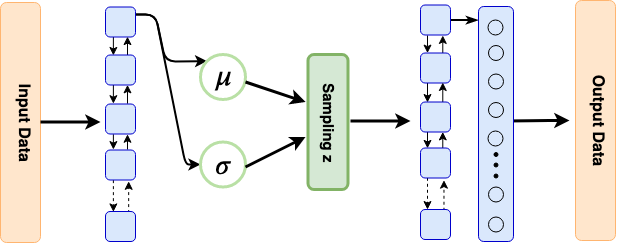}
        \centering\caption{Illustration of layout of bidirectional LSTM VAE built in this paper. The orange layers are the input and output. The blue squares represent the reading of sequences in the bidirectional LSTM layer. Both readings are then merged with a certain merge mode (summation or concatenation). The parameters are then estimated from the output of the bidirectional LSTM layer, to create the latent vector which is light green in the figure. The decoder consists of another bidirectional LSTM and it is followed by a dense layer which is the same size as the original data.}
    \end{figure}

\subsection{Changes in Loss Function of VAE}
The loss function of the VAE is comprised of a reconstruction loss and a variational regularization term.
The loss function is essential to help the decoder reconstruct a version of the data from the embeddings as close as possible to the input data. Binary cross-entropy is usually used as the reconstruction loss, however, we devised an F1-score based loss function to use as the reconstruction loss and compared its performance to binary cross-entropy. 

\[L=\mathbb{E}_{z\sim q}[\log p(x\mid z)]-D_{KL}(q(z\mid x)\mid \mid p(z))\]

    \begin{figure}[H]
        \centering\includegraphics[width=.75\linewidth]{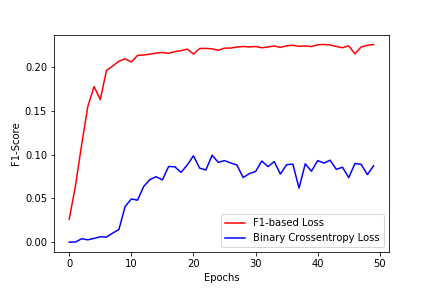}
        \centering\caption{Comparison of F1-loss (red) and Cross-entropy (blue) during training.}
    \end{figure}

Recent studies \cite{RN33} have been emphasizing the importance of the regularization term in the loss function and the role it plays in creating useful latent representations. We have explored beta-VAE \cite{RN34}, and different ways of introducing it.

\subsection{Assessment of Representation}
To assess the representation created by the VAE, 2-dimensional embeddings were created by the best VAE and compared to a lower dimension version of the data created using principal component analysis (PCA) \cite{RN16}.
Kmeans clustering was applied to both low-dimensional versions, and its results were compared to the known cancer types of each profile using Normalized Mututal Information (NMI) as a measure. The number of clusters created was 32 since there are 32 different cancer types.

The embeddings were also used in a given classification task and their performance was compared to results when the raw data was used in the same task.

\section{Results}
To assess the reconstruction ability of the VAE, we performed a 80/20 \% training/validation split, and used the micro F1 score. 
The MLP-VAE currently has a better performance than the bidirectional LSTM VAE, however, the preliminary results obtained by the bidirectional LSTM are promising and under investigation for future work.

The test F1-Score after a 5-fold cross-validation obtained with the MLP-VAE is 20.4\% and 17.1\% with the Bi-LSTM. Our analysis has shown that the size of the latent space does not play a big role in the reconstruction ability of the VAE as long it is above 8. Table \ref{table:latent-spaces} shows the validation F1 after a 5-fold cross-validation, found with different latent space sizes.

\begin{table}[H]
  \caption{Effect of Latent Space Size on VAE reconstruction}
  \label{table:latent-spaces}
  \centering
  \begin{tabular}{cccccccc}
    \toprule
    \textbf{Latent Space Size} & 2 & 8 & 32 & 64 & 128 & 265 & 512 \\
    \midrule
    \textbf{Validation F1 Score \%} & 15.68 & 20.55 & 20.88 & 20.41 & 20.50 & 20.18 & 20.21\\
    \bottomrule
  \end{tabular}
\end{table}

\begin{figure}[H]
  \centering
  \includegraphics[width=\linewidth]{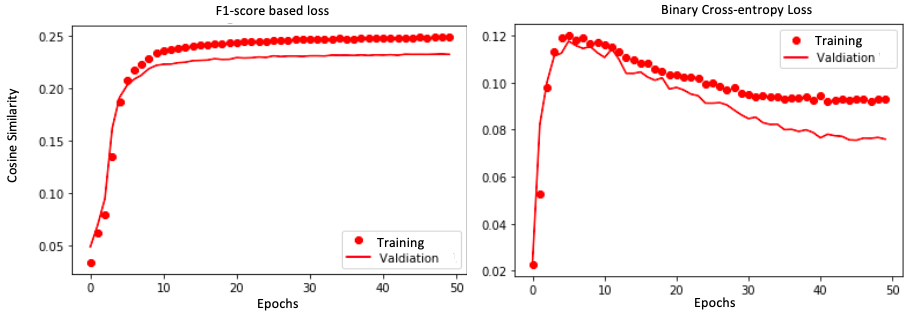}
  \caption{Comparison of F1-loss (left) and Cross-entropy (right) during training.}
  \label{figure:vae-training}
  \end{figure}

We found that the use of the F1-score based loss helps the VAE reconstruct this form of data much better than the binary cross-entropy loss. To assess this, we measure the cosine similarity. Figure \ref{figure:vae-training} shows the performance of the VAE while training with the two losses. We have also found that introducing the regularization term using a warm-up function and increasing its weight with each epoch gives the best performance.

\bigskip

\subsection{Visualization}
After performing Kmeans with 32 clusters on the embeddings from the best VAE and the lower dimension of the data created by PCA, plots with reference to the clusters were created, and they can be seen in Figure 2.
The NMI for the clusters created in the VAE embeddings was 21\%, and for the version created by PCA it was 11\% showing that the VAE embeddings are a better representation. We also note here that the data is binary, making it inherently hard to represent with PCA.

\begin{figure}[H]
  \centering
   \includegraphics[width=\linewidth]{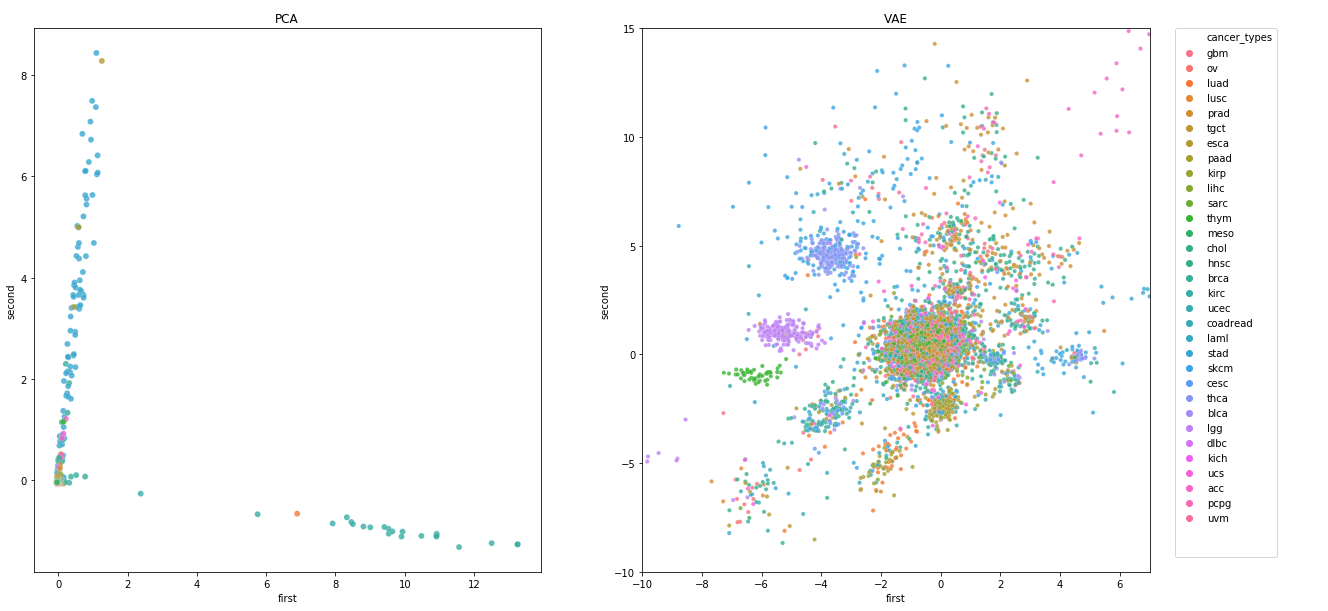}
  \caption{Comparison of projections from PCA (left) and Flatsomatic (right) for visualization.}
  \label{figure:kmeans-vis}
  \end{figure}

\section{Current Application \& Conclusion}

We have used Flatsomatic in a semi-supervised setting to predict drug response from the GDSC \cite{yang_genomics_2013} dataset. We show (Table \ref{table:drug-response}) that the performances are similar to the raw data. The advantage of working with smaller spaces is that it enables the use of other relevant features, for example clinical features, or additional abstract representations of further multi-omic profiles, all of which should be explored further.

The work done in this paper has shown that there is a great potential for the use of VAEs in creating utilizable lower dimension representations of somatic profiles. The VAE embeddings performed better than PCA for a clustering task, and performed equally well to the raw high dimensional data for a classification task.

\begin{table}[H]
  \caption{Predicting drug response with somatic profile}
  \label{table:drug-response}
  \centering
  \begin{tabular}{ccccc}
    \toprule
    Data source & Dimensionality & Precision & Recall & \textbf{F1-score} \\
    \midrule
    Mutations Counts       & 8298 & 0.732 & 0.614 & 0.667 \\
    Flatsomatic embeddings & 64 & 0.721 & 0.621 & 0.667 \\
    \bottomrule
  \end{tabular}
\end{table}

 Additional areas that would be suitable for further exploration would include the building and optimizing of other neural network architectures, in addition to applying the FlatSomatic embeddings to a semi-supervised task to increase classification performance.

\newpage

\bibliography{bib}

\begin{thebibliography}{18}
\providecommand{\natexlab}[1]{#1}
\providecommand{\url}[1]{\texttt{#1}}
\expandafter\ifx\csname urlstyle\endcsname\relax
  \providecommand{\doi}[1]{doi: #1}\else
  \providecommand{\doi}{doi: \begingroup \urlstyle{rm}\Url}\fi

\bibitem[Collisson et~al.(2019)Collisson, Bailey, Chang, and
  Biankin]{precisionpanc}
Eric~A. Collisson, Peter Bailey, David~K. Chang, and Andrew~V. Biankin.
\newblock Molecular subtypes of pancreatic cancer.
\newblock \emph{Nature Reviews Gastroenterology and Hepatology}, 16:\penalty0
  207--220, 2019.

\bibitem[Kumar-Sinha and Chinnaiyan(2018)]{precisiononcology}
Chandan Kumar-Sinha and Arul~M. Chinnaiyan.
\newblock Precision oncology in the age of integrative genomics.
\newblock \emph{Nature Biotechnology}, 36:\penalty0 46--60, 2018.

\bibitem[Dubourg-Felonneau et~al.(2018)Dubourg-Felonneau, Cannings, Cotter,
  Thompson, Patel, Cassidy, and Clifford]{prior}
Geoffroy Dubourg-Felonneau, Timothy~I. Cannings, Fergal Cotter, Hannah
  Thompson, Nirmesh Patel, John~W. Cassidy, and Harry~W. Clifford.
\newblock A framework for implementing machine learning on omics data.
\newblock \emph{ArXiv}, abs/1811.10455, 2018.

\bibitem[Tokheim et~al.(2016)Tokheim, Papadopoulos, Kinzler, Vogelstein, and
  Karchin]{drivers}
Collin~J Tokheim, Nickolas Papadopoulos, Kenneth~W. Kinzler, Bert Vogelstein,
  and Rachel Karchin.
\newblock Evaluating the evaluation of cancer driver genes.
\newblock \emph{Proceedings of the National Academy of Sciences of the United
  States of America}, 113 50:\penalty0 14330--14335, 2016.

\bibitem[Krupp et~al.(2010)Krupp, Maass, Marquardt, Staib, Bauer, K{\"o}nig,
  Biesterfeld, Galle, Tresch, and Teufel]{pathways}
Markus Krupp, Thorsten Maass, Jens~U. Marquardt, Frank Staib, Tobias Bauer,
  Rainer K{\"o}nig, Stefan Biesterfeld, Peter~R. Galle, Achim Tresch, and
  Andreas Teufel.
\newblock The functional cancer map: A systems-level synopsis of genetic
  deregulation in cancer, 2010.

\bibitem[Collins(2007)]{TCGA}
Asha~S Collins.
\newblock The cancer genome atlas (tcga) pilot project, 2007.

\bibitem[Bindal et~al.(2011)Bindal, Forbes, Beare, Gunasekaran, Leung, Kok,
  Jia, Bamford, Cole, Ward, Teague, Stratton, Campbell, and Futreal]{CCLP}
Nidhi Bindal, Simon~A. Forbes, David Beare, Prasad Gunasekaran, Kenric Leung,
  Chai~Yin Kok, Mingming Jia, Sally Bamford, Charlotte Cole, Sari Ward, Jon~W.
  Teague, Michael~R. Stratton, Peter~J. Campbell, and Andrew Futreal.
\newblock Cosmic: the catalogue of somatic mutations in cancer, 2011.

\bibitem[Bengio et~al.(2013)Bengio, Courville, and Vincent]{RN27}
Y.~Bengio, A.~Courville, and P.~Vincent.
\newblock Representation learning: A review and new perspectives.
\newblock \emph{IEEE Transactions on Pattern Analysis and Machine
  Intelligence}, 35\penalty0 (8):\penalty0 1798--1828, 2013.
\newblock ISSN 0162-8828.
\newblock \doi{10.1109/TPAMI.2013.50}.

\bibitem[Kingma and Welling(2013)]{RN15}
Diederik~P. Kingma and Max Welling.
\newblock \emph{Auto-Encoding Variational Bayes}, volume abs/1312.6114.
\newblock 2013.

\bibitem[Chollet et~al.(2015)]{RN40}
Francois Chollet et~al.
\newblock Keras.
\newblock 2015.
\newblock URL \url{https://keras.io}.

\bibitem[The Cancer Genome Atlas~Research et~al.(2013)The Cancer Genome
  Atlas~Research, Chang, Creighton, Davis, Donehower, Drummond, Wheeler, Ally,
  Balasundaram, Birol, Butterfield, Chu, Chuah, Chun, Dhalla, Guin, Hirst,
  Hirst, Holt, Jones, Lee, Li, Marra, Mayo, Moore, Mungall, Robertson, Schein,
  Sipahimalani, Tam, Thiessen, Varhol, Beroukhim, Bhatt, Brooks, Cherniack,
  Freeman, Gabriel, Helman, Jung, Meyerson, Ojesina, Pedamallu, Saksena,
  Schumacher, Tabak, Zack, Lander, Bristow, Hadjipanayis, Haseley,
  Kucherlapati, Lee, Lee, Luquette, Mahadeshwar, Pantazi, Parfenov, Park,
  Protopopov, Ren, Santoso, Seidman, Seth, Song, Tang, Xi, Xu, Yang, Zeng,
  Auman, Balu, Buda, Fan, Hoadley, Jones, Meng, Mieczkowski, Parker, Perou,
  Roach, Shi, Silva, Tan, Veluvolu, Waring, Wilkerson, Wu, Zhao, Bodenheimer,
  Hayes, Hoyle, Jeffreys, Mose, Simons, Soloway, Baylin, Berman, Bootwalla,
  Danilova, et~al.]{RN30}
Network The Cancer Genome Atlas~Research, Kyle Chang, Chad~J. Creighton, Caleb
  Davis, Lawrence Donehower, Jennifer Drummond, David Wheeler, Adrian Ally,
  Miruna Balasundaram, Inanc Birol, Yaron S.~N. Butterfield, Andy Chu, Eric
  Chuah, Hye-Jung~E. Chun, Noreen Dhalla, Ranabir Guin, Martin Hirst, Carrie
  Hirst, Robert~A. Holt, Steven J.~M. Jones, Darlene Lee, Haiyan~I. Li,
  Marco~A. Marra, Michael Mayo, Richard~A. Moore, Andrew~J. Mungall, A.~Gordon
  Robertson, Jacqueline~E. Schein, Payal Sipahimalani, Angela Tam, Nina
  Thiessen, Richard~J. Varhol, Rameen Beroukhim, Ami~S. Bhatt, Angela~N.
  Brooks, Andrew~D. Cherniack, Samuel~S. Freeman, Stacey~B. Gabriel, Elena
  Helman, Joonil Jung, Matthew Meyerson, Akinyemi~I. Ojesina, Chandra~Sekhar
  Pedamallu, Gordon Saksena, Steven~E. Schumacher, Barbara Tabak, Travis Zack,
  Eric~S. Lander, Christopher~A. Bristow, Angela Hadjipanayis, Psalm Haseley,
  Raju Kucherlapati, Semin Lee, Eunjung Lee, Lovelace~J. Luquette, Harshad~S.
  Mahadeshwar, Angeliki Pantazi, Michael Parfenov, Peter~J. Park, Alexei
  Protopopov, Xiaojia Ren, Netty Santoso, Jonathan Seidman, Sahil Seth, Xingzhi
  Song, Jiabin Tang, Ruibin Xi, Andrew~W. Xu, Lixing Yang, Dong Zeng, J.~Todd
  Auman, Saianand Balu, Elizabeth Buda, Cheng Fan, Katherine~A. Hoadley,
  Corbin~D. Jones, Shaowu Meng, Piotr~A. Mieczkowski, Joel~S. Parker,
  Charles~M. Perou, Jeffrey Roach, Yan Shi, Grace~O. Silva, Donghui Tan,
  Umadevi Veluvolu, Scot Waring, Matthew~D. Wilkerson, Junyuan Wu, Wei Zhao,
  Tom Bodenheimer, D.~Neil Hayes, Alan~P. Hoyle, Stuart~R. Jeffreys, Lisle~E.
  Mose, Janae~V. Simons, Mathew~G. Soloway, Stephen~B. Baylin, Benjamin~P.
  Berman, Moiz~S. Bootwalla, Ludmila Danilova, et~al.
\newblock The cancer genome atlas pan-cancer analysis project.
\newblock \emph{Nature Genetics}, 45:\penalty0 1113, 2013.
\newblock \doi{10.1038/ng.2764}.
\newblock URL \url{https://doi.org/10.1038/ng.2764}.

\bibitem[Ioffe and Szegedy(2015)]{RN39}
Sergey Ioffe and Christian Szegedy.
\newblock Batch normalization: Accelerating deep network training by reducing
  internal covariate shift.
\newblock \emph{arXiv preprint arXiv:1502.03167}, 2015.

\bibitem[He et~al.(2015)He, Zhang, Ren, and Sun]{LRELU}
Kaiming He, Xiangyu Zhang, Shaoqing Ren, and Jian Sun.
\newblock Delving deep into rectifiers: Surpassing human-level performance on
  imagenet classification, 2015.

\bibitem[Schuster and Paliwal(1997)]{Schuster1997BidirectionalRN}
Mike Schuster and Kuldip~K. Paliwal.
\newblock Bidirectional recurrent neural networks.
\newblock \emph{IEEE Trans. Signal Processing}, 45:\penalty0 2673--2681, 1997.

\bibitem[Alemi et~al.(2017)Alemi, Poole, Fischer, Dillon, Saurous, and
  Murphy]{RN33}
Alexander~A Alemi, Ben Poole, Ian Fischer, Joshua~V Dillon, Rif~A Saurous, and
  Kevin Murphy.
\newblock Fixing a broken elbo.
\newblock \emph{arXiv preprint arXiv:1711.00464}, 2017.

\bibitem[Higgins et~al.(2017)Higgins, Matthey, Pal, Burgess, Glorot, Botvinick,
  Mohamed, and Lerchner]{RN34}
Irina Higgins, Lo'efc Matthey, Arka Pal, Christopher Burgess, Xavier Glorot,
  Matthew~M Botvinick, Shakir Mohamed, and Alexander Lerchner.
\newblock \emph{beta-VAE: Learning Basic Visual Concepts with a Constrained
  Variational Framework}.
\newblock ICLR. 2017.

\bibitem[Wold et~al.(1987)Wold, Esbensen, and Geladi]{RN16}
Svante Wold, Kim Esbensen, and Paul Geladi.
\newblock Principal component analysis.
\newblock \emph{Chemometrics and Intelligent Laboratory Systems}, 2\penalty0
  (1):\penalty0 37--52, 1987.
\newblock ISSN 0169-7439.
\newblock \doi{https://doi.org/10.1016/0169-7439(87)80084-9}.
\newblock URL
  \url{http://www.sciencedirect.com/science/article/pii/0169743987800849}.

\bibitem[Yang et~al.()Yang, Soares, Greninger, Edelman, Lightfoot, Forbes,
  Bindal, Beare, Smith, Thompson, Ramaswamy, Futreal, Haber, Stratton, Benes,
  McDermott, and Garnett]{yang_genomics_2013}
W.~Yang, J.~Soares, P.~Greninger, E.~J. Edelman, H.~Lightfoot, S.~Forbes,
  N.~Bindal, D.~Beare, J.~A. Smith, I.~R. Thompson, S.~Ramaswamy, P.~A.
  Futreal, D.~A. Haber, M.~R. Stratton, C.~Benes, U.~McDermott, and M.~J.
  Garnett.
\newblock Genomics of {{Drug Sensitivity}} in {{Cancer}} ({{GDSC}}): A resource
  for therapeutic biomarker discovery in cancer cells.
\newblock 41:\penalty0 D955--61.
\newblock ISSN 0305-1048.
\newblock \doi{10.1093/nar/gks1111}.
\newblock URL \url{https://europepmc.org/articles/PMC3531057/}.

\end{thebibliography}

\end{document}